\documentclass[aps,pre, twocolumn, groupedaddress]{revtex4-1}
\usepackage{url}
\usepackage[utf8]{inputenc}
\usepackage{lipsum}
\usepackage{mathtools}
\usepackage{graphicx}
\usepackage{amsmath}    
\usepackage{amssymb}
\usepackage{bm}
\usepackage{hyperref}
\usepackage{latexsym}
\usepackage{verbatim}
\usepackage{color}
\usepackage{float}
\usepackage[caption=false]{subfig}
\usepackage{epstopdf}

\preprint{}

\usepackage{hyperref}

\begin{document}

\title{Effect of Weak Non-Conservative Dynamics on Pattern Formation in Scalar Active Matter}

\author{Sameer Kumar}
\email[]{sameerk@iitk.ac.in}
\affiliation{Department of Physics, Indian Institute of Technology Kanpur, Kanpur, India - 208016}

\date{\today}

\begin{abstract}

Biological systems such as bacteria and cells undergo growth or degradation, resulting in weak violations of mass conservation. We investigate how such weak non-conservative dynamics affect phase separation in scalar active matter by incorporating a reaction term into a minimal continuum model. Through numerical simulations and linear stability analysis, we show that even weak non-conservative reactions arrest coarsening and stabilize nonequilibrium microphase-separated states. With increasing activity, the system undergoes a morphological transition from interconnected labyrinthine patterns to worm-like structures and eventually to isolated droplets. Quantitative analysis of the correlation function and static structure factor reveals a well-defined steady-state characteristic length. Qualitative analysis of the resulting phases shows that the non-conservative reaction primarily promotes microphase separation and enhances local hexagonal ordering, while activity predominantly controls the domain morphology. Our results demonstrate that weak violations of mass conservation fundamentally alter the nonlinear coarsening dynamics of active phase separation and provide a minimal framework for understanding pattern formation in related systems.
\end{abstract}
\maketitle

\section{Introduction}
Active matter comprises systems whose constituent particles continuously consume energy from their surroundings and dissipate it to generate self-propelled motion, thereby driving the system intrinsically far from thermodynamic equilibrium \cite{MarchettiRMP2013, CatesARCMP2015, RamaswamyJStatMech2017}. Such systems are ubiquitous in nature and span a broad range of length scales, from microscopic systems including bacterial colonies \cite{Shapiro1995Significances, Harshey2003Bacterial, DombrowskiPRL2004, LiuPRL2019MXanthus}, epithelial cell monolayers \cite{FriedlNatPhys2009, Saw2017}, and cytoskeletal filaments \cite{DoostmohammadiNatComm2016}, to macroscopic systems such as bird flocks \cite{TonerPRL1995}, fish schools \cite{VicsekPRL1995}, etc. Unlike passive colloidal particles, whose assembly is dictated by equilibrium thermodynamics and external forcing, active particles continuously consume energy from their surroundings to generate self-propelled motion. This intrinsic nonequilibrium drive enables spontaneous self-organization and gives rise to a rich variety of collective phenomena, including motility-induced phase separation (MIPS), giant number fluctuations, traveling bands, and complex spatiotemporal patterns \cite{FilyPRL2012, TonerAoP2005, RamaswamySimhaToner2003, NarayanScience2007, VicsekPRL1995, DoostmohammadiNatComm2016, SameerPRL2025, JenaPRE2026}.

Owing to their continuous energy consumption and particle birth–death processes, many active systems, such as bacterial colonies, exhibit a rich variety of nonequilibrium patterns ranging from concentric rings to labyrinthine structures and droplet states \cite{BudreneNature1995, KawasakiJTBio1997Modeling, RamaswamySimhaToner2003, CatesPNAS2010, CatesARCMP2015}.

Among continuum descriptions of scalar active matter with a conserved order parameter, Active Model B (AMB) provides one of the simplest theoretical frameworks for studying active phase separation. It extends the Cahn--Hilliard equation through a nonequilibrium contribution to the chemical potential while retaining mass conservation, and generically undergoes coarsening toward bulk phase separation \cite{WittkowskiNatComm2014, PattanayakPRE2021, YadavPRE2025}. 

Many active systems encountered in nature, however, are not strictly conservative. Micro-organism (e.g. bacterial colonies, cells, etc.) continuously undergo particle birth and death, whereas synthetic active materials (e.g. active colloidal supersessions) may exchange matter with their surroundings through growth, degradation, or chemical reactions \cite{CatesPNAS2010, CuratoloNatPhy2020BactGrowth, BerryPNAS2015PhaseSepar, HubermanJCP2019}. Such processes introduce weak violations of mass conservation that can significantly influence the long-time evolution of phase-separated states. Recent studies adopted the coupled Cahn--Hilliard dynamics to chemical reactions and demonstrated the emergence of a variety of nonequilibrium structures, including microphase-separated and Turing-like patterns \cite{GlotzerPRL1995Reaction, ZwickerpRE2015Suppression, CatesPNAS2010, ChatelainNJP2011, Toffenetti2026Active, GibaudJPhys2009, MondalSM2025,  HupePRR2026, JenaSciRep2025SpatioTemporal}.

Despite these advances, a detailed study of the influence of {\em weak} non-conservative reaction dynamics on the late-time morphology and pattern selection in active phase separation is underexplored. In particular, how {\em weak} violations of mass conservation modify the nonlinear evolution and steady-state morphologies of scalar active systems has received comparatively little attention.

In this work, we investigate how weak non-conservative dynamics influence phase ordering in scalar active matter. To this end, we study a weakly non-conservative extension of Active Model B \cite{WittkowskiNatComm2014} by incorporating a reaction term, characterized by the coefficient $\gamma$, into the density evolution equation while retaining the standard active contribution to the chemical potential governed by the activity parameter $\zeta$. Through extensive numerical simulations and stability analysis, we demonstrate that even weak violations of mass conservation arrest coarsening and stabilize nonequilibrium microphase-separated states. Depending on the activity strength, the system exhibits a sequence of morphological transitions from interconnected labyrinthine domains to worm-like structures and eventually to isolated droplet phases. Structural characterization by calculating correlation functions, structure factors reveals that the systems selects an intrinsic length scale of activity strengths and reaction coefficient. Further the cluster statistics, average circularity of the isolated clusters, droplet-size distributions, and the hexatic bond-orientational order parameter reveals that the reaction term primarily promotes microphase separation and enhances local hexagonal ordering, whereas activity predominantly controls the domain morphology. To elucidate the origin of these nonequilibrium states, we complement our numerical simulations with a linear stability analysis, which distinguishes the roles of the reaction term and active chemical potential in the onset of phase separation. Our results demonstrate that weak violations of mass conservation fundamentally modify the nonlinear dynamics of active phase separation and provide a minimal framework for understanding pattern formation in active systems with non-conserved particle density.
 
\begin{figure}      
{\includegraphics[width=0.75\linewidth ]{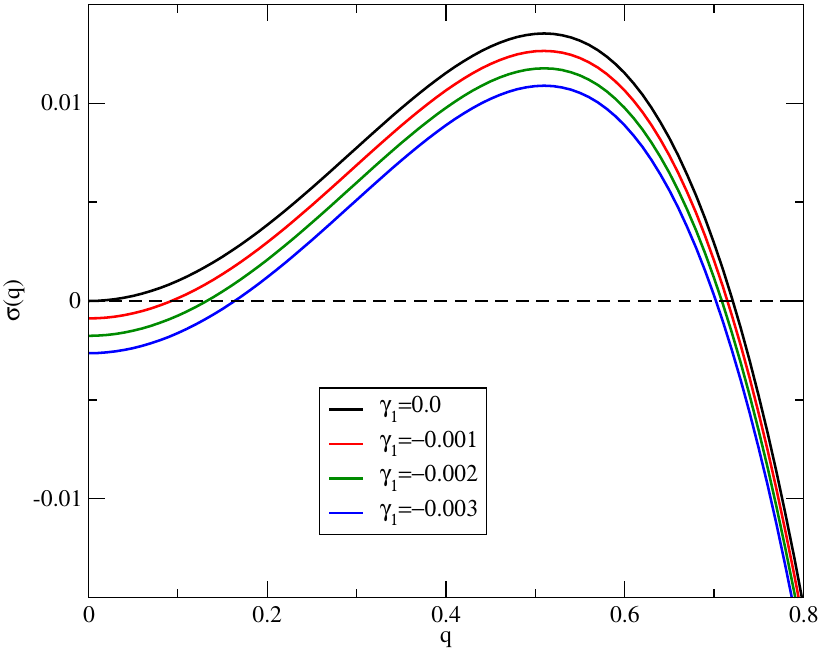}} 
\caption{ Growth rate $\sigma({\bf q})$ vs $q$ for different growth rate $\gamma$. }
\label{fig:1}
\end{figure}

\section{Model} 
We construct a coarse-grained  scaler order parameter field, $\phi({\bf r},t)$ that represent the density of the particles in a system. The ordered phase is described by the Landau free energy functional, $\mathcal{F}(\phi)=\int d^2 \bm{r} \left[A_\phi \left( -\frac{\phi^2}{2}+\frac{\phi^4}{4}\right)+ K_\phi \vert \nabla \phi \vert^2 \right],$ the first term in $\mathcal{F}(\phi)$ accounts for the bulk energy of the system whereas the second term penalize any gradient in the order parameter, $\phi({\bf r},t)$. 
The system evolves according to equation of motion given by,
\begin{equation}
    \centerline{$\frac{\partial \phi}{\partial t}=- \Gamma_\phi \bm{\nabla} \cdot \bm{J} + \gamma \phi \left(1-\frac{\phi}{\phi_c}\right) \left(1+\frac{\phi}{\phi_c}\right)$}
    \label{eq:1}
\end{equation}

The first term in Eq. \ref{eq:1} describe the current $\bm{J}({\bf r},t)$ defined as $\bm{J}= -\bm{\nabla} \mu + \sqrt{2\Gamma_\phi}{\bm\eta}
$ where $\mu$ is  the chemical potential given by,   $\mu({\bf r},t) = \frac{\delta \mathcal{F}}{\delta \phi} + \mu_{active}= -A_\phi \phi(1-\phi^2)- K_\phi \bm{\nabla}^2\phi + \zeta \vert \bm{\nabla} \phi \vert^2$. The presence of active part, $\mu_{active}=\zeta \vert \bm{\nabla} \phi \vert^2$ breaks the time reversal symmetry and makes the model ``Active Model B " and $\zeta$ controls the activity strength. Second term in Eq. \ref{eq:1} is the weak non-conservative reaction term  and its strength is controlled by the reaction term with the coefficient $\gamma$ and $\phi_c$ is some constant to make the system stable.  We keep the reaction coefficient, $\gamma \le 0$ such that the model converses to the standard Active Model B when $\gamma=0$ and it has a weak {\em sink} like term when $\gamma<0$. The noise term in $\bm{J}$ is the Gaussian white noise with $\langle \bm{\eta}({\bm r},t)\rangle=0$ and $\langle \bm{\eta}_i(t) \bm{\eta}_j(t^{'})\rangle=2 \Gamma_\phi \delta_{ij}\delta(t-t^{'})$.

\begin{figure*}      
{\includegraphics[width=0.9\linewidth]{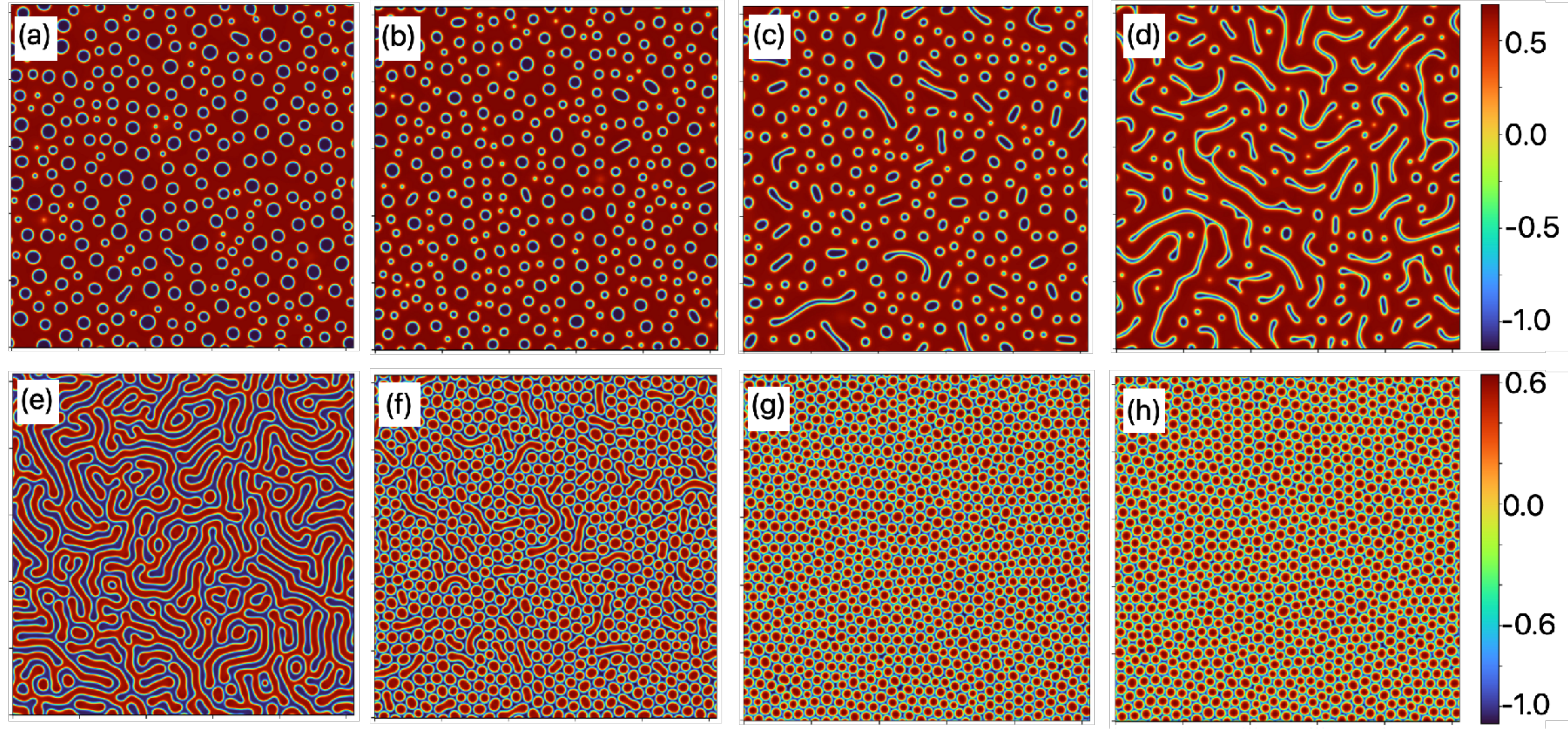}} 
\caption{ Snapshots from the simulation for the order parameter $\phi({\bf r},t)$ represented by the color map without the reaction term i.e., $\gamma=0.0$ (top panel) and with reaction coefficient  $\gamma=-0.001$ (bottom panel) for different strengths of activity,  $\zeta=-1.0$ (a,e), $-2.0$ (b,f), $-3.0$ (c,g) and $-4.0$ (d,h). Snapshots show the full simulation domain, $512 \times 512$.}
\label{fig:3}
\end{figure*}

\section{Stability analysis}
To check if the reaction term and/or the activity least to unstable modes we do the linear stability analysis of the system. We start by writing the dynamical equation for the phase field $\phi({\bf r},t)$,
\begin{equation}
    \centerline{$\partial_t \phi = \Gamma_\phi \nabla^2(-\phi+\phi^3-\nabla^2\phi+\zeta \vert \nabla \phi \vert^2)+\gamma\phi\left(1-\frac{\phi^2}{\phi_c^2}\right),$}
    \label{eq:3}
\end{equation}

we assume the homogeneous steady state is given by $\phi({\bf r},t)=\phi_0$, where $\phi_0$ is uniform. We put this in Eq \ref{eq:3} and get the homogeneous steady states,   $\phi_0=0$ and $\phi_0=\pm \phi_c$. 
We perturb the homogeneous steady state such that $\phi({\bf r},t)= \phi_0+\delta \phi({\bf r},t)$, with $\delta \phi \ll 1.$. Hence the  chemical potential in the linear order is given by, 
$$\delta \mu = (-1+3\phi_0^2)\delta \phi -\nabla^2\delta \phi.$$ Here the term $\zeta \vert \nabla \phi \vert^2$ would not survive in the linear order. Further the linearized reaction term is given by, $\delta R= \gamma (1-\frac{3\phi_0^2}{\phi_c^2})\delta \phi$. 
We substitute the above values in Eq. \ref{eq:3} and get linearized equation,
$$ \partial_t \delta \phi = -\Gamma_\phi(-1+3\phi_0^2)\nabla^2\delta \phi - \Gamma_\phi \nabla^4 \delta \phi +\gamma \left(1- \frac{3\phi_0^2}{\phi_c^2} \right)\delta \phi$$

We assume normal modes are defined as, $\delta \phi = \delta \phi_{\bf q}e^{\sigma({\bf q})+\iota {\bf q \cdot r}}$, and then  applying the Fourier Transform to linearized equation we get the dispersion relation,


\begin{equation}
    \centerline{$\sigma({\bf q})  = \Gamma_\phi(1-3\phi_0^2)q^2- \Gamma_\phi q^4  +\gamma \left(1- \frac{3\phi_0^2}{\phi_c^2}\right)$}
    \label{eq:2}
\end{equation}

We plot the growth rate, $\sigma({\bf q})$ on fig. \ref{fig:1} and find that the reaction coefficient $\gamma$ enters the dispersion relation as a q-independent constant, producing a vertical shift of the growth-rate curve without altering the position of its maximum. Consequently, the most unstable wave number $q^{*}$, and hence the characteristic wavelength selected during the early stage of phase separation, remain unchanged. Furthermore, the active contribution does not appear in the linearized dynamics because the active chemical potential, $\mu^{\textit{act}} =\zeta \vert \nabla \phi \vert^2$, is quadratic in the order-parameter gradients and therefore contributes only at nonlinear order. Thus, neither the reaction term nor the activity modifies the wavelength selected during the initial stage of phase separation, although the reaction coefficient changes the growth rate and the unstable band of wave numbers.

\begin{figure*}      
{\includegraphics[width=0.75\linewidth]{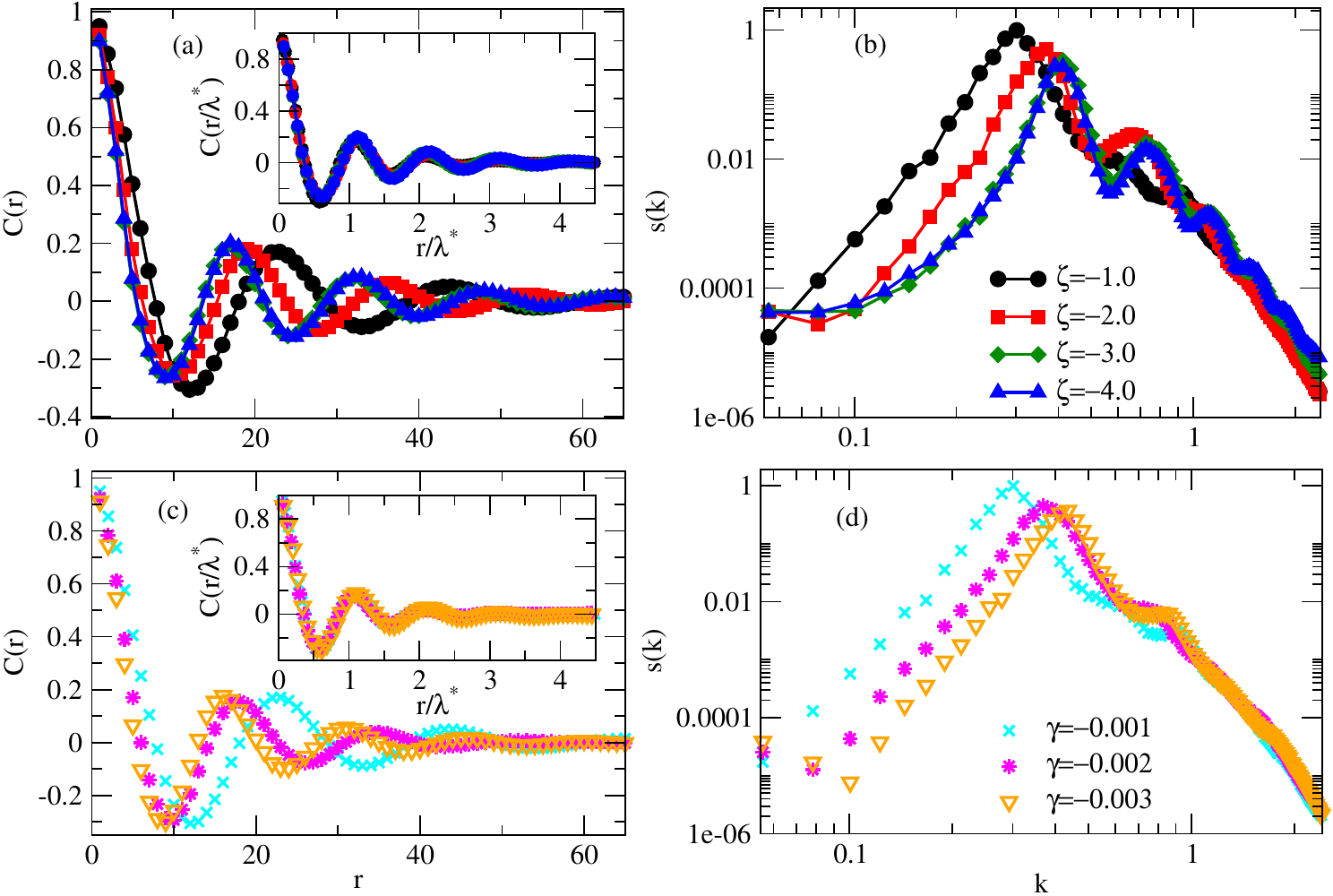}} 
\caption{  Correlation function $C(r)$ vs distance $r$ (a,c), scaled correlation function $C(r/\lambda^*)$ vs scaled distance $r/\lambda^*$ (a,c)insets. Structure factor $S({\bf k})$ vs wavenumber ${\bf k}$ (b,c). Top panel are the plots for reaction coefficient $\gamma=-0.001$ and different values of $\zeta$ (shows as different line colors with filled symbols). Bottom panel shows the data for $\zeta=-1.0$ and different values of reaction coefficient $\gamma$ represented by open symbols with different colors. Position of the first peak ($k^*$) in  $S({\bf k})$ vs  ${\bf k}$ gives us the characteristic  length scale defined as, $\lambda^*=2\pi/k^*$.}
\label{fig:5}
\end{figure*}
\section{Results}

We numerically integrate eq. (\ref{eq:1}) for different parameters with periodic boundary condition  for a square lattice ($L \times L$) and system size $L=512$. Some parameters are chosen to be constant and have the values as follows : $\Gamma_\phi=0.2, \ \bm{\eta}=0.0001\hat{\bm{e}}, \  A_\phi=1, \ K_\phi=1$ and $\phi_c=2.0$. We start the simulation with initial uniform state given by  $\phi_0= 2\rho-1$, where $\rho $ is the density of the particles in the system. We take, $\rho=0.7$ (or $\phi_0=0.4$) i.e. above critical configurations for all our simulation results. \\

We investigate the effect of the non-conservative reaction coefficient, $\gamma$ on the late-time morphologies of the system. Throughout this work, we consider $\gamma \leq 0$ and compare the results with the conserved case, $\gamma=0$ corresponding to the standard Active Model B (AMB). We shows late-time snapshots of the density order parameter $\phi({\bf r},t)$ for the conserved system $\gamma=0$ in Fig.\ref{fig:3}(a-d) and for the weakly non-conserved system with $\gamma=-0.001$ in Fig.\ref{fig:3}(e-h) for different values of the activity coefficient, $\zeta$.  The results demonstrate that the non-conservative reaction coefficient qualitatively change the activity-driven morphological evolution and broadly separates the system into two distinct regimes. In the conserved limit ( when $\gamma=0$), increasing the magnitude of the negative activity from $\zeta=-1.0$ to $-4.0$ drives a systematic structural change from a droplet morphology (Fig.\ref{fig:3}(a)) to worm-like structures (Fig.\ref{fig:3}(b,c)), and eventually to the system go through the spinodal decomposition forming a labyrinth state (Fig. \ref{fig:3}(d)). This behavior is consistent with the characteristic late-time evolution of the standard Active Model B. In contrast, in the presence of a weak non-conservative reaction term (when $\gamma=-0.001$), system qualitatively reverses this morphological evolution. As shown in Fig.\ref{fig:3}(e-h), increasing the magnitude of the activity now drives the system from a {\em labyrinth} morphology (for $\zeta=-1.0$) through an intermediate {\em worm-like} state ($\zeta=-2.0$), and finally to a microphase-separated {\em droplet} state for $\zeta=-3.0$ and $-4.0$. These results demonstrate that even a weak non-conservative reaction profoundly modifies the late-time coarsening dynamics, suppressing the formation of connected labyrinthine domains and instead promoting the stabilization of isolated droplet morphologies. Now in the following subsections we show  detailed analysis of the observed phases in the presence of weal non-conservative reaction term:\\

\subsection{Distinct phases for non-conserve case}
The system is initialized with a homogeneous density, $\phi_0 = 0.4$, and evolves in the presence of the weak non-conservative reaction term, $\gamma = -0.001$, we observe three distinct late-time morphologies depending on the activity parameter, $\zeta$. For weak activity ($\zeta = -1.0$), the system evolves into an  arrested {\em labyrinth phase} characterized by a well-defined stripe spanning throughout the system, see Fig. \ref{fig:3}(e). Upon increasing the activity to $\zeta = -2.0$, the labyrinth pattern partially fragments into isolated domains, giving rise to a  mixed lamellar-droplet or {\em worm-like phase}, where interconnected stripes coexist with nearly circular droplets, see Fig. \ref{fig:3}(f). For stronger activity (when $\zeta = -3.0$ and $-4.0$), the system undergoes complete { microphase separation}, consisting of  polydispersed high-density droplets embedded in a low-density background. We call this state as the {\em droplet phase}, see Fig. \ref{fig:3}(g,h). Remarkably, these droplets exhibit hexatic order that we  explore in detail in later section.\\

We further investigate the structural properties of the steady states by calculating the equal-time two-point correlation function, $
C(r)=\langle \phi({\bf r})\phi({\bf r}+{\bf d}r)\rangle,$ 
where $\langle\cdots\rangle$ denotes an average over many late-time steady-state configurations. The resulting correlation functions for different values of the activity parameter $\zeta$ are shown in Fig.~\ref{fig:5}(a). The damped oscillatory behavior of $C(r)$ is a clear signature of periodic microstructures with a well-defined characteristic length scale. In the subsequent sections, we further distinguish between the different morphologies using additional quantitative measures.
\begin{figure*}      
{\includegraphics[width=\linewidth]{ 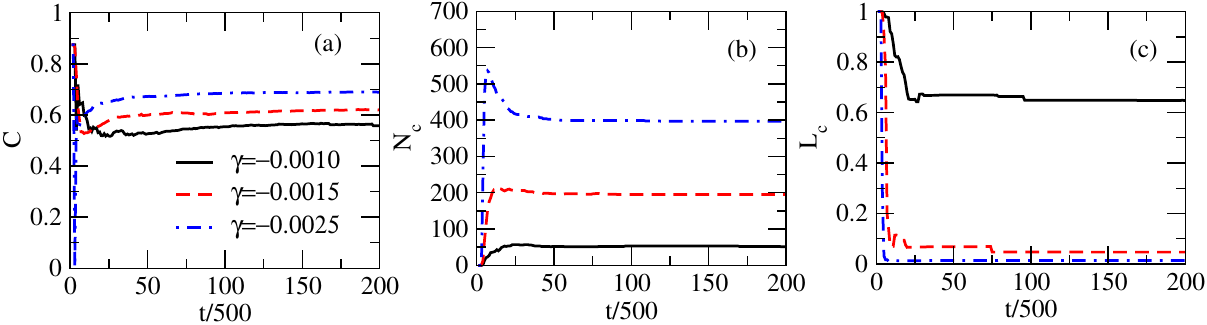}} 
\caption{ Time series plot of (a) circularity order parameter $C$, (b) number of isolated cluster $N_c$ and (c) average size of the largest connected domain $L_c$. Different lines shows the different values of reaction coefficients $\gamma$ and fixed activity strength $\zeta=-1.0$. }
\label{fig:8}
\end{figure*}
To characterize the spatial ordering in reciprocal space, we compute the static structure factor, $S({\bf k})$, for different values of $\zeta$, as shown in Fig.~\ref{fig:5}(b). The characteristic pattern size ($\lambda^{*}$) is obtained from the position of the first dominant peak, $k^*$, through the relation $\lambda^*=2\pi/k^*$. For $\zeta=-1.0$, we obtain $k^*\approx0.30$, with corresponding characteristic wavelength $\lambda^* \approx 20.9$. Similarly, for $\zeta = -2.0$, $-3.0$, and $-4.0$, the corresponding values are  $(k^*,\lambda^*) \simeq (0.37,17.0)$, $(0.41,15.3)$, and $(0.42,15.0)$, respectively. Thus, increasing the magnitude of the activity shifts the peak of the structure factor towards larger wave numbers, indicating a systematic reduction in the characteristic pattern size. Moreover, $\lambda^*$ remains unchanged at late times (data not shown), demonstrating that coarsening is arrested and the system evolves to a steady microphase-separated state with a well-defined intrinsic length scale.

To examine whether activity and the reaction coefficient modify only the characteristic length scale or also the underlying spatial organization, we plot the scaled correlation function, $C(r/\lambda^*)$, as a function of the scaled distance, $r/\lambda^*$, as shown in the insets of Fig.~\ref{fig:5}(a,c). Remarkably, the correlation functions for different values of $\zeta$ and $\gamma$ collapse onto a single master curve. This data collapse indicates that the two-point spatial correlations are governed by a common scaling form characterized by a single dominant length scale. Thus, variations in activity and reaction strength primarily rescale the characteristic length of the patterns while leaving their scaled two-point correlation structure largely unchanged. Importantly, this collapse does not imply identical morphologies, as changes in higher-order spatial organization can give rise to distinct domain morphologies while preserving the same two-point scaling form.

These observations collectively demonstrate that increasing the activity, in the presence of weak non-conservative dynamics, suppresses the formation of interconnected labyrinthine domains and stabilizes a microphase-separated droplet state with a well-defined intrinsic length scale. At large reaction rate, the droplets exhibit a  polydisperse distribution of the droplet size and structurally forms a hexatic order.

\subsection{Phase analysis}
To quantitatively characterize the different morphologies observed in the system, we calculate three different order parameters -  (a) the  largest cluster fraction defined as  $L_c=\frac{A_{\rm max}}{A_{\rm total}}$,
where $A_{\rm max}$ is the area of the largest connected cluster averaged over late-time configurations ($t=\mathcal{O}(10^5)$), and $A_{\rm total}=L^2$ is the total system area, (b) number of isolated clusters $N_c$, which counts the total number of disconnected domains with $\phi({\bf r},t)>0$ and (c) {\em circularity} order parameter defined as  $C=\frac{1}{N_c}\sum_{i=1}^{N_c}\frac{4\pi A_i}{P_i^2}$,
where $A_i$ and $P_i$ denote the area and perimeter, respectively, of the $i^{\rm th}$ connected cluster satisfying $\phi({\bf r},t)>0$; hence a value of $C=1$ corresponds to a perfect circular droplet, while smaller values indicate  elongated or irregular domains. Since circularity is a meaningful descriptor only for isolated compact domains, we evaluate the circularity order parameter only in the regime where the largest connected cluster no longer dominates the morphology. In our simulation results, this corresponds to $L_c <0.7$ for the present data set. We show for  the time series of these quantities for a chosen system parameters, i.e. activity strength $\zeta=-1.0$ and for different values of reaction coefficient $\gamma$ in fig. \ref{fig:8}. 

\begin{figure*}      
{\includegraphics[width=\linewidth]{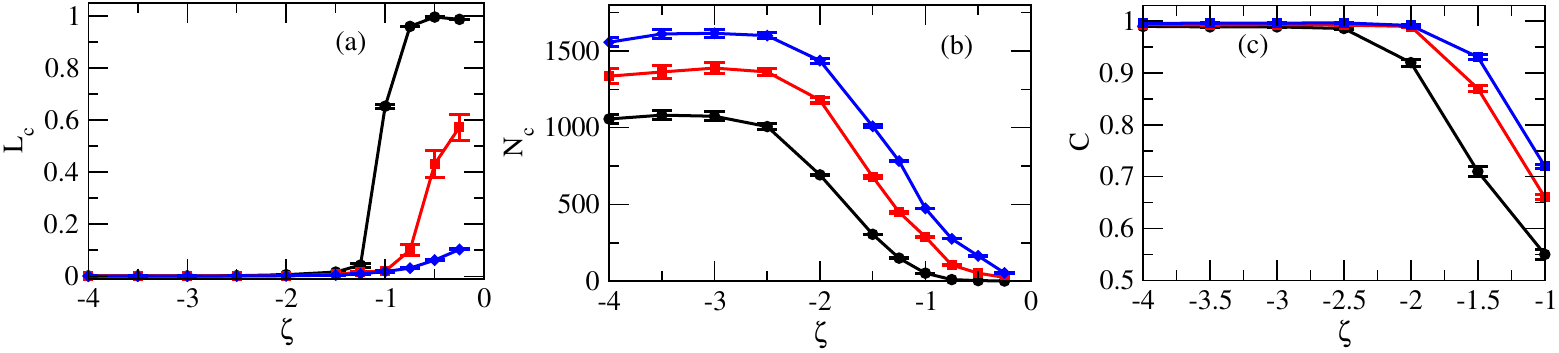}} 
\caption{ Plots of activity strength $\zeta$ vs. (a) largest cluster fraction $L_c$, (b) number of isolated clusters $N_c$ and (c) circularity order parameter $C$. All the plots are generated at late time for different growth rate $\gamma$ represent by different line color.}
\label{fig:6}
\end{figure*}

\begin{figure*}      
{\includegraphics[width=\linewidth]{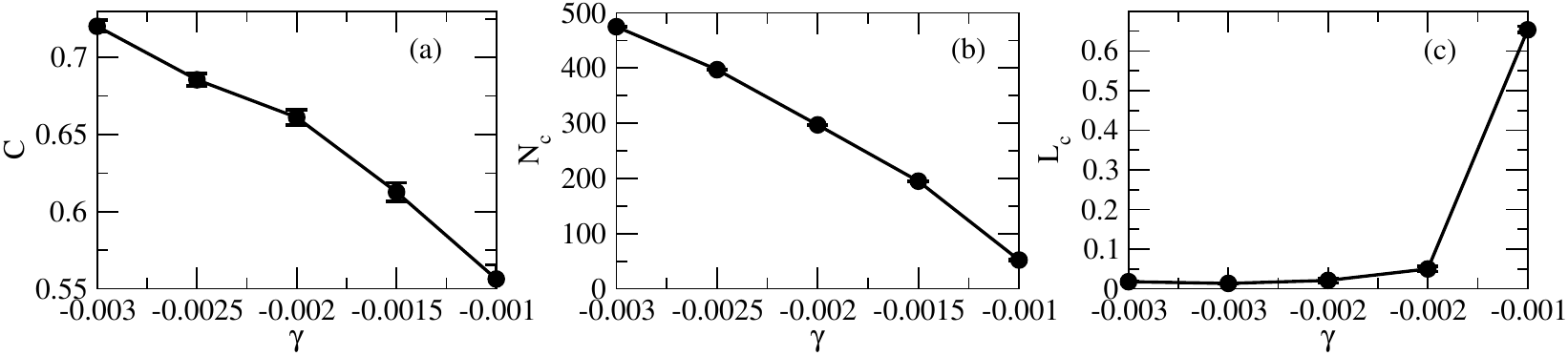}} 
\caption{ Plots of $\gamma$ vs. circularity $C$ (a), number of isolated clusters $N_c$ (b) and largest cluster fraction $L_c$ (c). All the plots are generated at late time ($t \sim \mathcal{O}(10^5)$)  for activity strength $\zeta=-1.0$.}
\label{fig:7}
\end{figure*}
Figure~\ref{fig:6}(a--d) summarizes the behavior of these quantities as functions of the activity parameter $\zeta$ for different values of the non-conservative reaction coefficient $\gamma$. Figures~\ref{fig:6}(a) and (c) show the average circularity, $C$, and the number of connected clusters, $N_c$, respectively. For all values of $\gamma$, both quantities remain nearly constant at large negative activity and decrease rapidly as $\zeta$ approaches zero. The large values of $C$ and $N_c$ at strong negative activity indicate the formation of numerous nearly circular, disconnected droplets, whereas weaker activity promotes domain coalescence, leading to elongated worm-like and labyrinthine structures.

Increasing the magnitude of the reaction coefficient shifts both $C$ and $N_c$ to larger values over the entire activity range, see Fig.~\ref{fig:7}, indicating that the non-conservative reaction suppresses domain coalescence and enhances fragmentation into isolated droplets. These results demonstrate that the activity parameter, $\zeta$, primarily controls the {\em type} of morphology, driving the crossover between labyrinthine and droplet phases, while the reaction coefficient, $\gamma$, governs the {\em degree of fragmentation} by controlling the number, size, and circularity of the isolated domains.

\begin{figure*}      
{\includegraphics[width=0.9\linewidth]{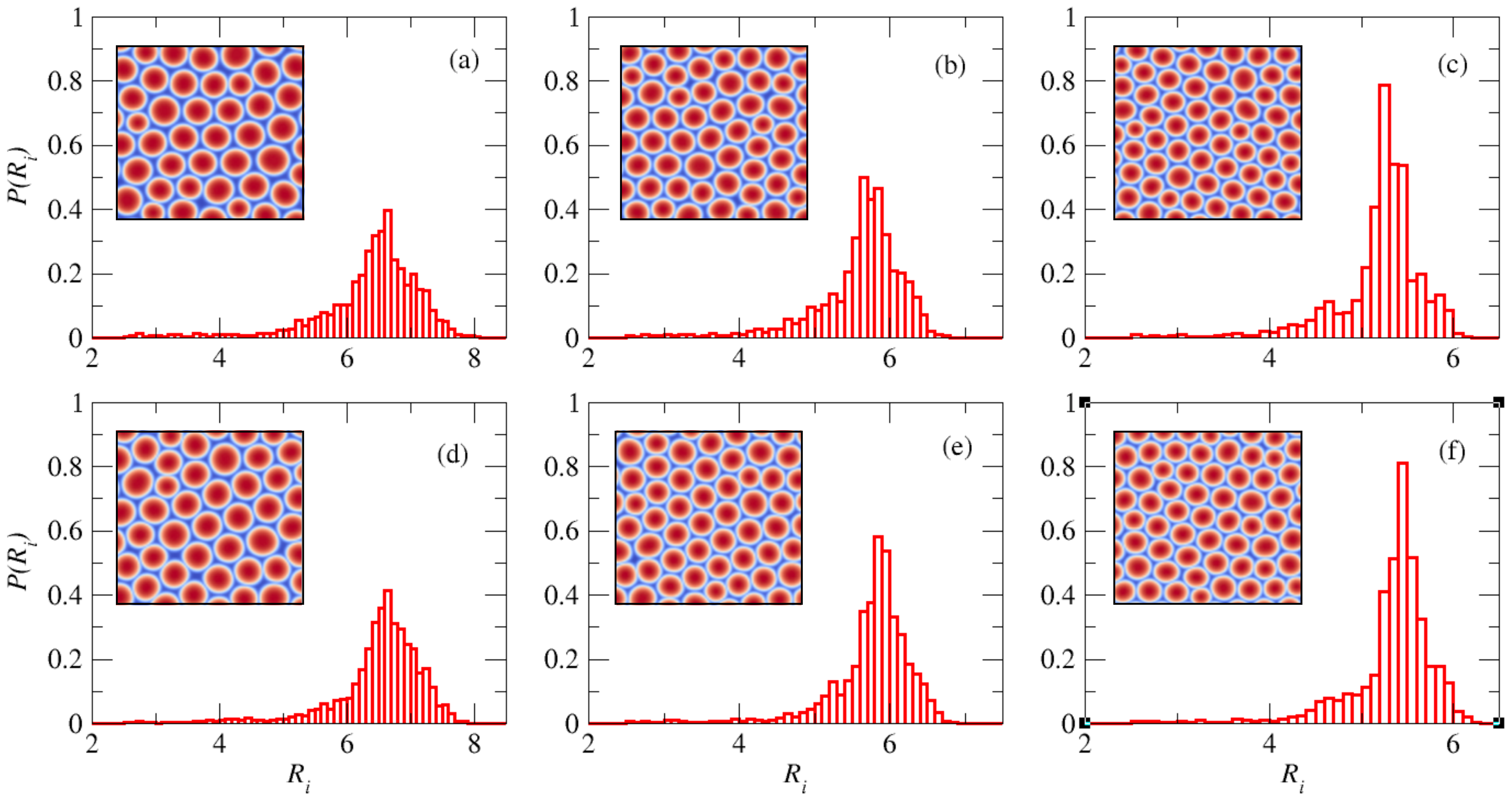}} 
\caption{ Probability distribution plot of equivalent radius $R_i=\sqrt{A_i/\pi}$ where $A_i$ is the area enclosed by $i^{\text{th}}$ cluster with circularity order parameter $C \simeq 1.$ for $\zeta=-3.0$ (top panel) and $\zeta=-4.0$  (bottom panel). Different columns are for reaction coefficient value, $\gamma=-0.001$ (a,d), $-0.002$ (b,e) and $-0.003$ (c,f). (Insets) snapshots from the simulation where we plot the contour $\phi({\bf r})=0$ in a subdomain of size $100 \times 100$. }
\label{fig:9}
\end{figure*}

We further investigated the structural properties of the microphase-separated states characterized by a circularity order parameter, $C \simeq 1$, indicating that the domains are nearly circular. To quantify the domain size, we computed the equivalent radius of each cluster, $R=\sqrt{\frac{A_i}{\pi}}$, where $A_i$ is the area enclosed by the $i^{\mathrm{th}}$ circular cluster (every isolated cluster is plotted as a contour of $\phi({\bf r},t)=0$). The corresponding probability distributions, $P(R)$, are shown in Fig.~\ref{fig:9} for activity parameters $\zeta=-3.0$ and $\zeta=-4.0$. The distributions are {\em unimodal} with a pronounced peak and exhibit a slight asymmetry rather than a perfect Gaussian form. As the magnitude of the reaction coefficient $\gamma$ is increased (from left to right), the peak of the distribution shifts systematically towards smaller radii and becomes progressively narrower for both values of the activity parameter. This trend indicates that the non-conserved reaction term enhances the fragmentation of the microphase-separated domains, thereby reducing their characteristic size while simultaneously suppressing the fluctuations in the domain-size distribution. These observations provide further evidence that the reaction term plays a dominant role in controlling the morphology and characteristic length scale of the steady-state microphase structure.

\begin{figure*}      
{\includegraphics[width=0.9\linewidth]{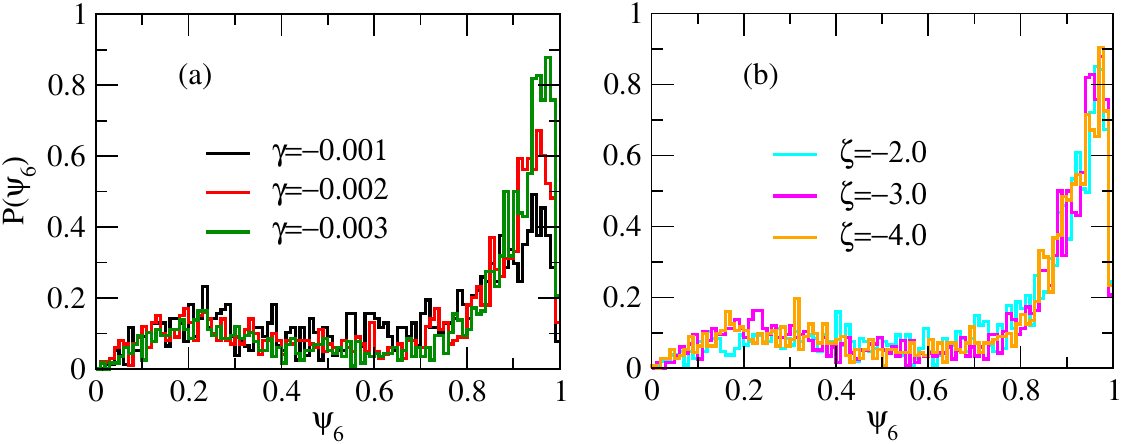}} 
\caption{ Probability distribution plot of $\psi_6$ for the clusters with circularity order parameter $C \simeq 1.0$ for (a) activity strength $\zeta=-1.0$ and different reaction coefficient $\gamma$. (b) For  reaction coefficient $\gamma=-0.003$ and different activity strength $\zeta$  (b). }
\label{fig:10}
\end{figure*}

Next, we quantify the local structural ordering within the droplet phase by calculating the hexatic bond-orientational order parameter, $\psi_6$, which characterizes the degree of local hexagonal packing (we can see the zoomed in snapshot where local hexatic order is evident in Fig. \ref{fig:9}-insets). The probability distribution, $P(\psi_6)$, is shown in Fig.~\ref{fig:10}(a) for a fixed activity strength, $\zeta=-1.0$, and different values of the non-conservative reaction coefficient, $\gamma_1$. In Fig.~\ref{fig:10}(b), we plot $P(\psi_6)$ for a fixed reaction coefficient, $\gamma_1=-0.003$, and different values of the activity parameter, $\zeta$.\\
For all the parameter values considered, the distribution exhibits a pronounced peak near $\psi_6=1$, indicating a strong tendency toward local hexagonal ordering within the droplet phase. Furthermore, for a fixed activity strength, the height of the peak near $\psi_6=1$ increases systematically with increasing magnitude of the reaction coefficient $\gamma_1$, demonstrating that the non-conservative reaction term not only promotes microphase separation but also enhances the local hexagonal close-packed (HCP) ordering of the droplets, as shown in Fig.~\ref{fig:10}(a). In contrast, varying the activity strength $\zeta$ has only a marginal effect on the distribution of $\psi_6$, indicating that activity primarily controls the characteristic length scale of the patterns without significantly altering the local HCP ordering, Fig.~\ref{fig:10}(b).

The system develops hexatic order through a coarsening process analogous to Ostwald ripening in a conserved system, where smaller droplets tend to disappear due to their higher chemical potential while larger droplets grow. In the present non-conserved system, however, this process cannot be identified with classical Ostwald ripening, since the order parameter is not conserved and its local value can be modified by the reaction term. Instead, the dynamics results from the competition between curvature-driven coarsening, active transport, and the non-conservative reaction, which ultimately arrests the growth at a finite characteristic droplet size (characterized by the characteristic length scale $\lambda^{*}$). As illustrated in the Supplementary Movie, the system evolves toward a state in which a large fraction of the droplets acquire comparable sizes. This size selection is evident from the equivalent-radius distributions, $P(R_i)$, which become progressively narrower with increasing reaction rate $\gamma$. The corresponding increase in hexatic order is reflected in the distribution of the local bond-orientational order parameter, $P(\psi_6)$, which develops an increasingly pronounced peak at large $\psi_6$, approaching unity. Thus, our results suggest that activity is responsible for sustaining the microphase-separated state, whereas the non-conservative reaction plays a crucial role in regulating the droplet-size distribution. The resulting size homogenization facilitates the packing of droplets into an ordered arrangement and thereby promotes the emergence of hexatic order.

\section{Discussion}

In this work, we have investigated the effect of weak non-conservative dynamics on phase ordering in scalar active matter by introducing a reaction term into the density evolution equation. Our results demonstrate that even weak violations of mass conservation qualitatively modify the nonlinear coarsening dynamics and the resulting steady-state morphologies. In the conserved limit ($\gamma=0$), corresponding to the standard Active Model B \cite{WittkowskiNatComm2014, CatesARCMP2015}, increasing the magnitude of the activity parameter drives the familiar sequence of droplet (binodal decomposition), worm-like, and labyrinthine (spinodal decomposition) morphologies. In contrast, the inclusion of a weak non-conservative reaction ($\gamma < 0$) arrests coarsening and stabilizes nonequilibrium microphase-separated states \cite{GlotzerPRL1995Reaction, CatesPNAS2010}. Depending on the activity strength, the system undergoes a morphological transition from labyrinthine structures to worm-like domains and eventually to isolated droplet phases.

The linear stability analysis provides insight into the origin of these distinct patterns. While the reaction term modifies the growth rate of unstable modes without altering the most unstable wavelength, the active contribution to the chemical potential enters only through the nonlinear dynamics. Consequently, the qualitative differences between the conserved and weakly non-conserved systems arise primarily from the nonlinear evolution following the initial instability rather than from changes in the instability itself. Physically, the weak reaction continuously counteracts mass redistribution between neighboring domains, thereby suppressing coarsening and stabilizing finite-sized structures with an intrinsic steady-state length scale. The linear stability analysis demonstrates that the reaction term only weakly modifies the initial instability, whereas the simulations reveal that its principal effect is manifested during the nonlinear coarsening stage, where the interplay between weak non-conservative dynamics and active interfacial transport qualitatively drives the late-time morphology.

A detailed structural characterization reveals distinct roles for the activity parameter and the reaction coefficient. Correlation function and structure factor analyses demonstrate the emergence of a well-defined intrinsic length scale associated with the arrested patterns. Furthermore, the collapse of the scaled correlation functions onto a universal master curve indicates that activity and reaction term primarily rescales the characteristic pattern size while preserving the normalized spatial organization \cite{BrayAdvPhys1994, TjhungPRX2018}. Quantitative analyses of the observed patterns consistently show that the reaction coefficient promotes microphase separation, suppresses domain coalescence, and enhances local hexagonal ordering, whereas activity predominantly governs the global morphology and characteristic wavelength of the patterns.

The present model is motivated by the observation that many natural and synthetic active systems are not strictly conservative. Biological systems continuously undergo particle birth, death, growth, and degradation \cite{CatesPNAS2010, ZwickerpRE2015Suppression, GelimsonPRL2015}, while synthetic active materials can exchange matter with their surroundings through chemical reactions or external reservoirs \cite{PalacciSci2013, BechingerRMP2016}. Such processes weakly violate mass conservation and are therefore expected to influence the long-time dynamics and morphology of phase-separated states. In this context, the present model provides a minimal continuum framework for investigating the interplay between activity and weak non-conservative dynamics.

These findings highlight non-conservative dynamics as a simple yet effective mechanism for controlling nonequilibrium self-organization in active matter and may provide useful guidance for understanding pattern formation in living systems and for designing synthetic active materials with tunable microstructures.\\

\section{acknowledgment}
I would like to thank Shradha Mishra for introducing me to the problem and for useful discussions. I am also grateful to Manas Khan for insightful discussions on a broader problem, which inspired me to explore active systems with non-conserved reactions. The support and the resources provided by PARAM Sanganak under the National Supercomputing Mission, Government of India at the Indian Institute of Technology, Kanpur are gratefully acknowledged.

\bibliographystyle{apsrev4-1}
\bibliography{references} 

@preamble{ " \newcommand{\noop}[1]{} " }

@article{CatesPNAS2010,
  author  = {Cates, M. E. and Marenduzzo, D. and Pagonabarraga, I. and Tailleur, J.},
  title   = {Arrested phase separation in reproducing bacteria creates a generic route to pattern formation},
  journal = {Proceedings of the National Academy of Sciences},
  year    = {2010},
  volume  = {107},
  number  = {26},
  pages   = {11715--11720},
  doi     = {10.1073/pnas.1001994107}
}

@article{FilyPRL2012,
  title = {Athermal Phase Separation of Self-Propelled Particles with No Alignment},
  author = {Fily, Yaouen and Marchetti, M. Cristina},
  journal = {Phys. Rev. Lett.},
  volume = {108},
  issue = {23},
  pages = {235702},
  numpages = {5},
  year = {2012},
  month = {Jun},
  publisher = {American Physical Society},
  doi = {10.1103/PhysRevLett.108.235702},
  url = {https://link.aps.org/doi/10.1103/PhysRevLett.108.235702}
}

@article{PattanayakPRE2021,
  title = {Ordering kinetics in the active model $B$},
  author = {Pattanayak, Sudipta and Mishra, Shradha and Puri, Sanjay},
  journal = {Phys. Rev. E},
  volume = {104},
  issue = {1},
  pages = {014606},
  numpages = {7},
  year = {2021},
  month = {Jul},
  publisher = {American Physical Society},
  doi = {10.1103/PhysRevE.104.014606},
  url = {https://link.aps.org/doi/10.1103/PhysRevE.104.014606}
}

@article{DombrowskiPRL2004,
author = {Dombrowski, Christopher and Cisneros, Luis and Chatkaew, Sunita and Goldstein, Raymond E. and Kessler, John O.},
doi = {10.1103/PhysRevLett.93.098103},
issn = {00319007},
journal = {Phys. Rev. Lett.},
number = {9},
pages = {2--5},
title = {{Self-concentration and large-scale coherence in bacterial dynamics}},
volume = {93},
year = {2004}
}

@article{TonerPRL1995,
  title = {Long-Range Order in a Two-Dimensional Dynamical $\mathrm{XY}$ Model: How Birds Fly Together},
  author = {Toner, John and Tu, Yuhai},
  journal = {Phys. Rev. Lett.},
  volume = {75},
  issue = {23},
  pages = {4326--4329},
  numpages = {0},
  year = {1995},
  month = {Dec},
  publisher = {American Physical Society},
  doi = {10.1103/PhysRevLett.75.4326},
  url = {https://link.aps.org/doi/10.1103/PhysRevLett.75.4326}
}

@article{VicsekPRL1995,
  title = {Novel Type of Phase Transition in a System of Self-Driven Particles},
  author = {Vicsek, Tam\'as and Czir\'ok, Andr\'as and Ben-Jacob, Eshel and Cohen, Inon and Shochet, Ofer},
  journal = {Phys. Rev. Lett.},
  volume = {75},
  issue = {6},
  pages = {1226--1229},
  numpages = {0},
  year = {1995},
  month = {Aug},
  publisher = {American Physical Society},
  doi = {10.1103/PhysRevLett.75.1226},
  url = {https://link.aps.org/doi/10.1103/PhysRevLett.75.1226}
}

@article {NarayanScience2007,
	author = {Narayan, Vijay and Ramaswamy, Sriram and Menon, Narayanan},
	title = {Long-Lived Giant Number Fluctuations in a Swarming Granular Nematic},
	volume = {317},
	number = {5834},
	pages = {105--108},
	year = {2007},
	doi = {10.1126/science.1140414},
	publisher = {American Association for the Advancement of Science},
	issn = {0036-8075},
	URL = {http://science.sciencemag.org/content/317/5834/105},
	journal = {Science}
}

@article{MarchettiRMP2013,
title = {Hydrodynamics of soft active matter},
author = {Marchetti, M. C. and Joanny, J. F. and Ramaswamy, S. and Liverpool, T. B. and Prost, J. and Rao, Madan and Simha, R. Aditi},
journal = {Review of Modern Physics},
volume = {85},
issue = {3},
pages = {1143--1189},
numpages = {0},
year = {2013},
month = {Jul},
publisher = {American Physical Society},
doi = {10.1103/RevModPhys.85.1143},
url = {https://link.aps.org/doi/10.1103/RevModPhys.85.1143}
}

@article{TonerAoP2005,
title = "Hydrodynamics and phases of flocks",
journal = "Annals of Physics",
volume = "318",
number = "1",
pages = "170 - 244",
year = "2005",
note = "Special Issue",
issn = "0003-4916",
doi = "https://doi.org/10.1016/j.aop.2005.04.011",
url = "http://www.sciencedirect.com/science/article/pii/S0003491605000540",
author = "John Toner and Yuhai Tu and Sriram Ramaswamy"
}

@article{CatesARCMP2015,
author = {Michael E. Cates and Julien Tailleur},
title = {Motility-Induced Phase Separation},
journal = {Annual Review of Condensed Matter Physics},
volume = {6},
number = {1},
pages = {219-244},
year = {2015},
doi = {10.1146/annurev-conmatphys-031214-014710},
URL = { https://doi.org/10.1146/annurev-conmatphys-031214-014710 }
}

@Article{DoostmohammadiNatComm2016,
author={Doostmohammadi, Amin
and Adamer, Michael F.
and Thampi, Sumesh P.
and Yeomans, Julia M.},
title={Stabilization of active matter by flow-vortex lattices and defect ordering},
journal={Nat. Comm.},
year={2016},
month={Feb},
day={03},
publisher={The Author(s) SN  -},
volume={7},
pages={10557 EP  -},
note={Article},
url={http://dx.doi.org/10.1038/ncomms10557}
}

@article{RamaswamyJStatMech2017,
  author={Sriram Ramaswamy},
  title={Active matter},
  journal={Journal of Statistical Mechanics: Theory and Experiment},
  volume={2017},
  number={5},
  pages={054002},
  url={http://stacks.iop.org/1742-5468/2017/i=5/a=054002},
  year={2017}
}

@article{BrayAdvPhys1994,
author = { A.J.   Bray },
title = {Theory of phase-ordering kinetics},
journal = {Advances in Physics},
volume = {43},
number = {3},
pages = {357-459},
year  = {1994},
publisher = {Taylor & Francis},
doi = {10.1080/00018739400101505},

URL = { 
        https://doi.org/10.1080/00018739400101505
    
},
eprint = { 
        https://doi.org/10.1080/00018739400101505
    
}

}

@Article{WittkowskiNatComm2014,
author={Wittkowski, Raphael
and Tiribocchi, Adriano
and Stenhammar, Joakim
and Allen, Rosalind J.
and Marenduzzo, Davide
and Cates, Michael E.},
title={Scalar ph4 field theory for active-particle phase separation},
journal={Nat. Comm.},
year={2014},
month={Jul},
day={10},
publisher={Nature Publishing Group, a division of Macmillan Publishers Limited. All Rights Reserved. SN  -},
volume={5},
pages={4351 EP  -},
note={Article},
url={http://dx.doi.org/10.1038/ncomms5351}
}

@article{BechingerRMP2016,
  title = {Active particles in complex and crowded environments},
  author = {Bechinger, Clemens and Di Leonardo, Roberto and L\"owen, Hartmut and Reichhardt, Charles and Volpe, Giorgio and Volpe, Giovanni},
  journal = {Review Modern Physics},
  volume = {88},
  issue = {4},
  pages = {045006},
  numpages = {50},
  year = {2016},
  month = {Nov},
  publisher = {American Physical Society},
  doi = {10.1103/RevModPhys.88.045006},
  url = {https://link.aps.org/doi/10.1103/RevModPhys.88.045006}
}

@article{Shapiro1995Significances,
  author  = {Shapiro, James A.},
  title   = {The significances of bacterial colony patterns},
  journal = {BioEssays},
  year    = {1995},
  volume  = {17},
  number  = {7},
  pages   = {597--607},
  doi     = {10.1002/bies.950170706}
}

@article{GlotzerPRL1995Reaction,
  author  = {Glotzer, Sharon C. and Di Marzio, Edmund A. and Muthukumar, M.},
  title   = {Reaction-Controlled Morphology of Phase-Separating Mixtures},
  journal = {Physical Review Letters},
  year    = {1995},
  volume  = {74},
  number  = {11},
  pages   = {2034--2037},
  doi     = {10.1103/PhysRevLett.74.2034}
}

@article{ZwickerpRE2015Suppression,
  author  = {Zwicker, David and Hyman, Anthony A. and J{\"u}licher, Frank},
  title   = {Suppression of Ostwald ripening in active emulsions},
  journal = {Physical Review E},
  year    = {2015},
  volume  = {92},
  number  = {1},
  pages   = {012317},
  doi     = {10.1103/PhysRevE.92.012317}
}

@misc{Toffenetti2026Active,
  author        = {Toffenetti, Davide and Nettuno, Beatrice and Weyer, Henrik and Frey, Erwin},
  title         = {Active Model {B}$^-$ from Mass-Conserving Reaction-Diffusion Systems},
  year          = {2026},
  eprint        = {2605.15903},
  archivePrefix = {arXiv},
  primaryClass  = {cond-mat.soft},
  doi           = {10.48550/arXiv.2605.15903}
}

@article{KawasakiJTBio1997Modeling,
  author  = {Kawasaki, K. and Mochizuki, A. and Matsushita, M. and Umeda, T. and Shigesada, N.},
  title   = {Modeling spatio-temporal patterns generated by {Bacillus} subtilis},
  journal = {Journal of Theoretical Biology},
  year    = {1997},
  volume  = {188},
  number  = {2},
  pages   = {177--185},
  doi     = {10.1006/jtbi.1997.0462}
}

@article{BudreneNature1995,
  author  = {Budrene, Elena O. and Berg, Howard C.},
  title   = {Dynamics of formation of symmetrical patterns by chemotactic bacteria},
  journal = {Nature},
  year    = {1995},
  volume  = {376},
  number  = {6535},
  pages   = {49--53},
  doi     = {10.1038/376049a0}
}

@article{Harshey2003Bacterial,
  author  = {Harshey, Rasika M.},
  title   = {Bacterial motility on a surface: Many ways to a common goal},
  journal = {Annual Review of Microbiology},
  year    = {2003},
  volume  = {57},
  number  = {1},
  pages   = {249--273},
  doi     = {10.1146/annurev.micro.57.030502.091014}
}

@article{RamaswamySimhaToner2003,
	doi = {10.1209/epl/i2003-00346-7},
	url = {https://doi.org/10.1209/epl/i2003-00346-7},
	year = 2003,
	month = {apr},
	publisher = {{IOP} Publishing},
	volume = {62},
	number = {2},
	pages = {196--202},
	author = {S Ramaswamy and R. Aditi Simha and J Toner},
	title = {Active nematics on a substrate: Giant number fluctuations and long-time tails},
	journal = {Europhysics Letters ({EPL})}
}

@article{SameerPRL2025,
  title = {Hydrodynamic Bend Instability of Motile Particles on a Substrate},
  author = {Kumar, Sameer and de Graaf Sousa, Niels and Doostmohammadi, Amin},
  journal = {Phys. Rev. Lett.},
  volume = {135},
  issue = {26},
  pages = {268302},
  numpages = {7},
  year = {2025},
  month = {Dec},
  publisher = {American Physical Society},
  doi = {10.1103/jfsd-sg1v},
  url = {https://link.aps.org/doi/10.1103/jfsd-sg1v}
}

@article{JenaSciRep2025SpatioTemporal,
  author  = {Jena, Pratikshya and Mishra, Shradha},
  title   = {Spatio-temporal patterns in growing bacterial suspensions},
  journal = {Scientific Reports},
  year    = {2025},
  volume  = {15},
  number  = {1},
  pages   = {30948},
  doi     = {10.1038/s41598-025-13297-5}
}

@article{YadavPRE2025,
  title = {Coarsening kinetics in active model $\mathrm{B}+$: Macroscale and microscale phase separation},
  author = {Yadav, Pradeep Kumar and Mishra, Shradha and Puri, Sanjay},
  journal = {Phys. Rev. E},
  volume = {112},
  issue = {3},
  pages = {035412},
  numpages = {11},
  year = {2025},
  month = {Sep},
  publisher = {American Physical Society},
  doi = {10.1103/kb9k-w7jr},
  url = {https://link.aps.org/doi/10.1103/kb9k-w7jr}
}

@article{JenaPRE2026,
  title = {Competing effect of disorder on phase separation in active systems},
  author = {Jena, Pratikshya and Dikshit, Shambhavi and Mishra, Shradha},
  journal = {Phys. Rev. E},
  volume = {113},
  issue = {1},
  pages = {014128},
  numpages = {14},
  year = {2026},
  month = {Jan},
  publisher = {American Physical Society},
  doi = {10.1103/j741-jhzw},
  url = {https://link.aps.org/doi/10.1103/j741-jhzw}
}

@article{PalacciSci2013,
  author  = {Palacci, Jeremie and Ab{\'e}cassis, Benjamin and Cottin-Bizonne, C{\'e}cile and Yake, Christophe and Yodh, Arjun G. and Bocquet, Lyd{\'e}ric},
  title   = {Living crystals of light-activated colloidal surfers},
  journal = {Science},
  year    = {2013},
  volume  = {339},
  number  = {6122},
  pages   = {936--940},
  doi     = {10.1126/science.1230020}
}

@article{GelimsonPRL2015,
  author  = {Gelimson, Anatoly and Golestanian, Ramin},
  title   = {Collective dynamics of dividing active microorganisms},
  journal = {Physical Review Letters},
  year    = {2015},
  volume  = {114},
  number  = {2},
  pages   = {028101},
  doi     = {10.1103/PhysRevLett.114.028101}
}

@article{TjhungPRX2018,
  author  = {Tjhung, E. and Nardini, C. and Cates, M. E.},
  title   = {Cluster phases and active foam in active {Model B+}},
  journal = {Physical Review X},
  year    = {2018},
  volume  = {8},
  number  = {3},
  pages   = {031080},
  doi     = {10.1103/PhysRevX.8.031080}
}

@article{HubermanJCP2019,
    author = {Huberman, B. A.},
    title = {Striations in chemical reactions},
    journal = {The Journal of Chemical Physics},
    volume = {65},
    number = {5},
    pages = {2013-2019},
    year = {1976},
    month = {09},
    issn = {0021-9606},
    doi = {10.1063/1.433272},
    url = {https://doi.org/10.1063/1.433272}
}

@article{ChatelainNJP2011,
  doi = {10.1088/1367-2630/13/11/115013},
  url = {https://doi.org/10.1088/1367-2630/13/11/115013},
  year = {2011},
  month = {nov},
  publisher = {IOP Publishing},
  volume = {13},
  number = {11},
  pages = {115013},
  author = {C Chatelain and B Berche and W Janke},
  title = {Monte Carlo study of phase transitions in the three-dimensional random-field Ising model},
  journal = {New Journal of Physics}
}

@article{CuratoloNatPhy2020BactGrowth,
  author  = {Curatolo, Agnese I. and Zhou, Nan and Zhao, Yongfeng and Liu, Chenli and Daerr, Adrian and Tailleur, Julien and Huang, Jian-Dong},
  title   = {Cooperative pattern formation in multi-component bacterial systems through reciprocal motility regulation},
  journal = {Nature Physics},
  year    = {2020},
  volume  = {16},
  number  = {11},
  pages   = {1152--1157},
  doi     = {10.1038/s41567-020-0964-z}
}

@article{HupePRR2026,
  title = {Phase separation in a mixture of proliferating and motile active matter},
  author = {Hupe, Lukas and Materska, Joanna M. and Zwicker, David and Golestanian, Ramin and Waclaw, Bartlomiej and Bittihn, Philip},
  journal = {Phys. Rev. Res.},
  volume = {8},
  issue = {2},
  pages = {L022012},
  numpages = {7},
  year = {2026},
  month = {Apr},
  publisher = {American Physical Society},
  doi = {10.1103/9pns-h5ll},
  url = {https://link.aps.org/doi/10.1103/9pns-h5ll}
}

@article{MondalSM2025,
  author  = {Mondal, Sayantan and Das, Prasenjit},
  title   = {Phase separation in active binary mixtures with chemical reaction},
  journal = {Soft Matter},
  year    = {2025},
  volume  = {21},
  number  = {20},
  pages   = {4093--4100},
  doi     = {10.1039/D5SM00263J}
}

@article{GibaudJPhys2009,
  author  = {Gibaud, Thomas and Schurtenberger, Peter},
  title   = {A closer look at arrested spinodal decomposition in protein solutions},
  journal = {Journal of Physics: Condensed Matter},
  year    = {2009},
  volume  = {21},
  number  = {32},
  pages   = {322201},
  doi     = {10.1088/0953-8984/21/32/322201}
}

@article{BerryPNAS2015PhaseSepar,
  author  = {Berry, Joel and Weber, Stephanie C. and Vaidya, Nilesh and Haataja, Mikko and Brangwynne, Clifford P.},
  title   = {RNA transcription modulates phase transition-driven nuclear body assembly},
  journal = {Proceedings of the National Academy of Sciences},
  year    = {2015},
  volume  = {112},
  number  = {38},
  pages   = {E5237--E5245},
  doi     = {10.1073/pnas.1509317112}
}

@article{LiuPRL2019MXanthus,
  title = {Self-Driven Phase Transitions Drive Myxococcus xanthus Fruiting Body Formation},
  author = {Liu, Guannan and Patch, Adam and Bahar, Fatmag\"ul and Yllanes, David and Welch, Roy D. and Marchetti, M. Cristina and Thutupalli, Shashi and Shaevitz, Joshua W.},
  journal = {Phys. Rev. Lett.},
  volume = {122},
  issue = {24},
  pages = {248102},
  numpages = {6},
  year = {2019},
  month = {Jun},
  publisher = {American Physical Society},
  doi = {10.1103/PhysRevLett.122.248102},
  url = {https://link.aps.org/doi/10.1103/PhysRevLett.122.248102}
}

@article{Saw2017,
author = {Saw, Thuan Beng and Doostmohammadi, Amin and Nier, Vincent and Kocgozlu, Leyla and Thampi, Sumesh and Toyama, Yusuke and Marcq, Philippe and Lim, Chwee Teck and Yeomans, Julia M. and Ladoux, Benoit},
doi = {10.1038/nature21718},
issn = {14764687},
journal = {Nature},
number = {7649},
pages = {212--216},
publisher = {Nature Publishing Group},
title = {{Topological defects in epithelia govern cell death and extrusion}},
url = {http://dx.doi.org/10.1038/nature21718},
volume = {544},
year = {2017}
}

@ARTICLE{FriedlNatPhys2009,
author = {Friedl, P. and Gilmour, D.},
title = {Collective cell migration in morphogenesis, regeneration and cancer},
journal = {Nature Reviews Molecular Cell Biology},
year = {2009},
month = {july},
volume = {10},
pages = {445–457},
doi = {doi.org/10.1038/nrm2720},
surl = {https://doi.org/10.1038/nrm2720}
}
\end{document}